\documentclass[aip,reprint]{revtex4-1}
\usepackage{amsmath,amsthm,amssymb,bm}
\usepackage[dvips]{graphicx}

\begin{document}

\newcommand{\bi}[1]{\ensuremath{\boldsymbol{#1}}}

\title{Polaronic nature of a muonium-related paramagnetic center in SrTiO$_3$}

\author{T.~U.~Ito}
\email[]{ito.takashi15@jaea.go.jp}
\affiliation{Advanced Science Research Center, Japan Atomic Energy
 Agency, Tokai, Ibaraki 319-1195, Japan}
\author{W.~Higemoto}
\affiliation{Advanced Science Research Center, Japan Atomic Energy
 Agency, Tokai, Ibaraki 319-1195, Japan}
\affiliation{Department of Physics, Tokyo Institute of Technology, 
Meguro, Tokyo 152-8551, Japan}
\author{A.~Koda}
\author{K.~Shimomura} 
\affiliation{Institute of Materials Structure Science, High Energy
Accelerator Research Organization (KEK), Tsukuba,
Ibaraki 305-0801, Japan}

\date{\today}

\begin{abstract}
The hyperfine features and thermal stability of a muonium (Mu)-related
 paramagnetic center were investigated in the SrTiO$_3$
 perovskite titanate via muon spin rotation spectroscopy.
 The hyperfine coupling tensor of the paramagnetic center was
 found to have prominent dipolar characteristics, indicating that
 the electron spin density is dominantly distributed on a
 Ti site to form a small polaron near an ionized Mu$^+$ donor.
 Based on a hydrogen-Mu analogy, interstitial hydrogen is also expected to form
 such a polaronic center in the dilute doping limit. The small
 activation energy of 30(3)~meV found for the thermal dissociation of the
 Mu$^+$-polaron complex suggests that the strain energy required to distort the
 lattice is comparable to the electronic energy gained by localizing the
 electron.

\end{abstract}


\maketitle





Electron doping into an archetypal perovskite titanate, SrTiO$_3$,
results in intriguing physical properties such as
superconductivity\cite{schooley64} and ferromagnetism,\cite{rice14} and
can help form a two-dimensional (2D) electron gas system at the
LaAlO$_3$/SrTiO$_3$ heterointerface.\cite{ohtomo04,yoshimatsu08}
This makes electron-doped SrTiO$_3$ a
promising material for emerging oxide electronics. 
For many of its applications, it is crucial to determine whether excess electrons form
either free or trapped carriers. However, experimental studies on (La, Nb)-doped or
oxygen-deficient SrTiO$_3$ seem contradictory on this point.
Optical conductivity and dc transport measurements
showed that excess electrons behave as free carriers or large
polarons having band-like characteristics.\cite{mechelen08,devreese10} 
On the other hand, photoemission
spectroscopy (PES) studies revealed an incoherent in-gap state, which
can be associated with small polarons, and its
coexistence with a coherent delocalized state.\cite{ishida08} 
The Kondo effect observed in an electric field-effect-induced 2D
electron system in SrTiO$_3$ also suggests the coexistence of free and
trapped carriers.\cite{lee11} Although many theoretical studies have
been conducted to understand such puzzling
phenomena,\cite{janotti14,hao15,iwazaki14} the origin of the
dual electron behavior in $n$-type SrTiO$_3$ remains unclear.

Interstitial hydrogen (H$_i$) also serves as a shallow donor in
SrTiO$_3$.\cite{nakayama18}
PES measurements on a SrTiO$_3$ (001) {\it surface} exposed to hot
atomic hydrogen showed both in-gap and delocalized
states.\cite{dangelo12} This implies that small polarons can also exist
in H$_i$-doped SrTiO$_3$. However, little is known about the 
microscopic nature of hydrogen dopants and doped electrons in {\it bulk}
SrTiO$_3$ except for some insight provided by infrared (IR) and
polarized Raman scattering 
spectroscopies\cite{kapphan80,weber86,klauer94}
and positive muon spin rotation ($\mu^+$SR)
spectroscopy.\cite{spencer83,salman14}. 
Further experimental investigations are required to fully
elucidate the role of H$_i$ in {\it bulk} SrTiO$_3$ and
better understand the behavior of excess electrons
in $n$-type SrTiO$_3$.

Since hydrogen and muonium (Mu: a $\mu^+$-$e^-$ bound state) have 
almost the same reduced mass, the $\mu^+$SR technique has been
extensively used for the study of hydrogen-related defects in condensed
matter.\cite{spencer83,salman14,cox06,ito13,ito17,vilao15,shimomura15}
In this letter, we report a detailed $\mu^+$SR study in bulk SrTiO$_3$,
focusing on the electronic structure of a Mu-related paramagnetic defect
that results from muon implantation. Only two $\mu^+$SR studies on
SrTiO$_3$ have been published to date.\cite{spencer83,salman14} These
reported shallow Mu$^0$-like features with ionization energies of 
several tens of meV and hyperfine coupling parameters in
the MHz range.
Salman {\it et al.} performed measurements at 25~K in a zero applied
field (ZF) and transverse fields (TFs) below 12~mT and obtained a fully
anisotropic hyperfine coupling tensor for the Mu$^0$-like
center.\cite{salman14} However, the physical identity of the
paramagnetic center is far from being fully elucidated.
In this study, we carefully evaluated the hyperfine coupling tensor at 
1.7~K under TFs above 0.1~T. This tensor showed 
prominent dipolar characteristics, suggesting that the electron
spin density is dominantly distributed on a Ti site adjacent to an
interstitial muon (Mu$^+_i$) bound to an O$^{2-}$ ion.
Such a feature indicates that the small polaron can also form in
``Mu-doped'' bulk SrTiO$_3$ near an ionized Mu$^+_i$ donor.
Our temperature variation study revealed that the
Mu$^+_i$-polaron complex easily dissociates at moderate temperatures
with an activation energy of 30(3)~meV.
Similar small polarons bound to positively charged impurities may
explain the dual behavior of excess electrons observed in other
electron-doped SrTiO$_3$ systems.\cite{ishida08,lee11}



A nominally undoped SrTiO$_3$ single crystal of
10$\times$10$\times$0.5~mm$^3$ was obtained from Furuuchi Chemical Co.,
Japan. The single crystal was grown by the Verneuil method and cut along
the cubic (001) plane. It is worth mentioning that we did not actively
control the domain distribution in an antiferro-distortive (AFD) phase
below 105~K. Hence, we use cubic notation to specify crystallographic
directions for both cubic and pseudocubic-AFD phases.
$\mu^+$SR experiments were performed at Paul Scherrer Institut (PSI),
 Switzerland, using the GPS instrument\cite{amato17} and a
 spin-polarized surface muon beam in the TF configuration.
 The SrTiO$_3$ single crystal was mounted on
 a low-background sample holder with the (001) plane perpendicular to
 the muon incident direction. TFs up to 0.5~T were applied along the [001] direction.
 The asymmetry of the muon beta decay was monitored by the ``Up''
 and ``Down'' positron counters and recorded in the temperature range 1.7-300~K.

 \begin{figure}
\includegraphics[scale =0.48]{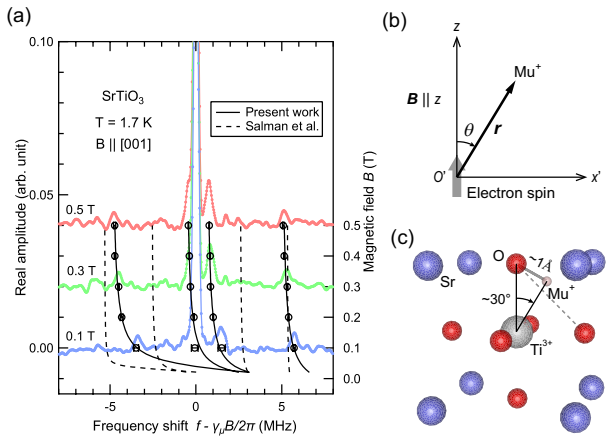}
  \caption{\label{fig1} (a) FT-$\mu^+$SR spectra at 1.7~K in TFs of 0.1,
  0.3, and 0.5~T applied along the [001] direction (small solid circles
  with lines). The left vertical axis and the horizontal axis correspond
  to the real Fourier amplitude and the frequency shift $f-\gamma_{\mu}B/2\pi$,
  respectively. The $B$ dependences of paramagnetic frequency
  shifts are also shown with large open circles. The right vertical axis
  represents $B$. The solid curves are the best fits to $f(B; \theta,
  A_{iso}, A_{dip})$ calculated from  eqs.~(\ref{eq1}) and
  (\ref{eq2}). The dashed curves are the high-field extension of
  paramagnetic frequency shifts calculated with
    hyperfine parameters in Ref.~[\cite{salman14}]. (b) Coordinate system
  ($x^{\prime},y^{\prime},z$) for the point-dipole model. (c) Schematic
  illustration of the Mu$^+_i$-bound small polaron in SrTiO$_3$.}
   \end{figure}
Figure~\ref{fig1}(a) shows the TF-$\mu^+$SR spectra in the frequency domain
(Fourier transform (FT) spectra) at 1.7~K for several TFs. The horizontal
axis corresponds to the frequency shift which is defined as
$f-\gamma_{\mu}B/2\pi$, where $f$ is the muon spin precession frequency,
$\gamma_{\mu}$(=2$\pi\times$135.53~MHz) is the muon
gyromagnetic ratio, and $B$ is the magnetic field.
Two pairs of satellite lines were clearly identified, indicating that a
paramagnetic electron is localized near the muon and the hyperfine
coupling between these particles is strongly anisotropic. The central
frequencies of the four paramagnetic lines, superimposed in Fig.~\ref{fig1}(a)
as functions of $B$, were obtained from time domain
fits to an oscillatory function composed of five exponentially-damped
cosines (one for a diamagnetic line at $f=\gamma_{\mu}B/2\pi$).
The largest principal value of the hyperfine coupling tensor was roughly
estimated to be in the order of 10$^1$~MHz from the splitting of the outer
pair. This is much smaller than the electron Zeeman frequency above
0.1~T ($>$2.8~GHz). Hence, the $z$ component of the electron spin
operator $\bm{s}$ becomes a good quantum number. Under such 
conditions, two lines in each pair simply correspond to
$s_z=\pm\frac{1}{2}$ states. If the hyperfine coupling is isotropic, as
in the case of atomic Mu$^0$, the electron spin creates a collinear
hyperfine field at the muon position and the splitting
should be symmetric in terms of the diamagnetic muon frequency $\gamma_{\mu}B/2\pi$.
This is clearly not the case in SrTiO$_3$, for which asymmetric shifts
were observed in both satellite pairs (Fig.~\ref{fig1}(a)). In addition, the
asymmetry is more remarkable at lower $B$ values. This strongly 
suggests that the hyperfine fields corresponding to the four paramagnetic
lines are noncollinear to $\bm{B}$ and therefore a dipolar contribution is dominant in the
hyperfine coupling.\cite{vilao15} Such a situation is expected to occur
when an electron is localized primarily at a single cation in close proximity to a
Mu$^+_i$ donor to form a polaronic center.\cite{cox06} 

To verify this hypothesis quantitatively, we analyzed the field dependences of 
the paramagnetic shifts using the following spin Hamiltonian $\mathcal{H}$, 
\begin{equation}
\mathcal{H}/h =
 \bm{\nu_e}\cdot\bm{s}-\bm{\nu_{\mu}}\cdot\bm{I}+(A_{iso}+\bm{\mathsf{A_{dip}}})\bm{s}\cdot\bm{I}, \label{eq1}
\end{equation}
where $\bm{\nu_e}=|g| \mu_{B}\bm{B}/h$ is the electron Zeeman frequency, 
$g$($\sim$2) is the electron $g$-factor; 
$\bm{\nu_{\mu}}=\gamma_{\mu}\bm{B}/2\pi$ is the muon Zeeman
frequency, $\bi{I}$ is the muon spin operator, $A_{iso}$
is the isotropic hyperfine coupling constant due to the Fermi
contact interaction, and $\bm{\mathsf{A_{dip}}}$ is the
dipolar hyperfine coupling tensor.
Here we adopt a point-dipole model to describe $\bm{\mathsf{A_{dip}}}$
in detail, where the electron spin is
localized at the origin, as depicted in Fig. 1(b), and creates a dipolar field at the muon
location $\bm{r}$. In the TF configuration, $\mu^+$SR is insensitive to
the crystal rotation around the field axis along the $z$
direction in the laboratory frame ($x,y,z$).
To simplify the expression for $\bm{\mathsf{A_{dip}}}$, another coordinate system
($x^{\prime},y^{\prime},z$) for the point-dipole model was selected so
that the $x^{\prime}$-$z$ plane contains $\bm{r}$.
In this frame, $\bm{\mathsf{A_{dip}}}$ is expressed as follows, 
\begin{align}
\bm{\mathsf{A_{dip}}} &=\frac{A_{dip}}{2}
    \begin{pmatrix}
     3\sin^2\theta -1 &0 &3\sin\theta \cos\theta \\
     0 &-1 &0\\
     3\sin\theta \cos\theta &0 &3\cos^2\theta -1
    \end{pmatrix},\label{eq2}\\
 A_{dip}&=\frac{\mu_0\gamma_{\mu}|g| \mu_{eff}}{4\pi^2r^3},\label{eq3} 
\end{align}
where $\theta$ is the angle between $\bm{B}$ and $\bm{r}$, $\mu_{eff}$
is the effective magnetic moment for the localized electron,
and $r$(=$|\bm{r}|$) is the electron-muon distance.
With this expression a pair of paramagnetic frequencies can be calculated for
each $\theta$.
A global fit to the calculated $f(B; \theta, A_{iso}, A_{dip})$ was
performed for the inner and outer frequency pairs with shared $A_{iso}$ and
$A_{dip}$. Solid curves in Fig.~\ref{fig1}(a) represent the best
fit with $A_{iso}=1.4(3)$~MHz, $A_{dip}=15.5(2)$~MHz,
$\theta=33(1)^{\circ}$ or 180$^{\circ}-33(1)^{\circ}$
for the outer pair, and $\theta=62(1)^{\circ}$ or
180$^{\circ}-62(1)^{\circ}$ for the inner
pair. This point-dipole model reproduced the field-dependent
asymmetric feature of the paramagnetic lines in $\bm{B}\parallel [001]$ well.

The hyperfine parameters ($A_{dip}\gg A_{iso}$) suggest that the spin
density distribution is considerably compact and
the center of gravity of the distribution is displaced from the muon
position by an atomic-scale length. This is similar to the case wherein 
a small polaron is trapped near an
ionized donor. In titanates, the small polaron is often found in the form of
a nominally Ti$^{3+}$ ($d^1$) ion accompanied by a
significant lattice distortion.\cite{ishida08,setvin14}
Thus, we assume
that an electron spin and a Mu$^+$ donor are respectively located on a Ti
site and an interstitial site near an oxygen ion to form an OMu$^-$
group in the same unit cell.
Based on the atomic configuration around H$^+_i$ reported from
first-principles calculations,\cite{iwazaki14} Mu$^+_i$ 
is expected to locate at $\theta=29^{\circ}$ on a
TiO$_2$ plane near an O-O edge of an oxygen octahedron (Fig.~\ref{fig1}(c)).
This agrees with $\theta$ for the outer paramagnetic
pair and also with $\theta$ for the inner pair after a $90^{\circ}$-rotation
of the system in the TiO$_2$ plane. Therefore, the two sets of satellite
pairs can be assigned to the identical paramagnetic centers in different
AFD domains. The Mu$^+_i$ site shown in Fig.~\ref{fig1}(c) is also
consistent with the H$^+_i$ site determined from IR and
polarized Raman scattering measurements in the cubic and pseudocubic-AFD
phases.\cite{kapphan80,weber86,klauer94}
When the Ti-H distance is considered as $r=2.2$\AA~in eq.~(\ref{eq3})
based on a theoretical structure,~\cite{iwazaki14} $\mu_{eff}/\mu_B$
becomes 0.33(1), which is of order unity and therefore consistent with the characteristics of
the nominally-Ti$^{3+}$ small polaron. 
Thus, we conclude that the identity of the Mu$^0$-like
paramagnetic center is a Mu$^+_i$-bound small
polaron. From the Mu-H analogy, H$_i$ in SrTiO$_3$ is also
expected to act as an auto-ionized shallow donor that can trap one 
excess electron in the form of the nominally-Ti$^{3+}$ small polaron, at
least in the dilute doping limit.
It is worth mentioning that a similar Mu$^+_i$-polaron complex was
reportedly found in rutile TiO$_2$.\cite{vilao15,shimomura15}
 However, there seems to be ambiguity as to the degree of electron
 localization since the hyperfine parameters for this state
($\sim$1~MHz) are too small to ascribe to a Ti$^{3+}$ ion in close
proximity to Mu$^+_i$.

 \begin{figure}
 \includegraphics[scale =0.48]{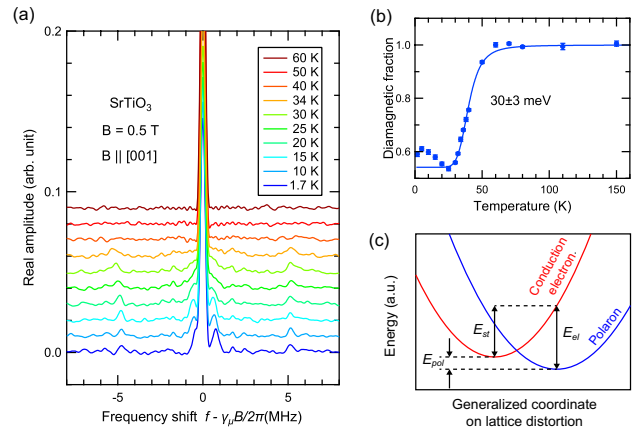}
  \caption{\label{fig2} (a) Temperature evolution of FT-$\mu^+$SR spectra in 
  a TF of 0.5~T applied along the [001] direction. (b) Temperature
  dependence of the diamagnetic fraction. The solid curve represents the
  best fit to the ionization model for data points above 25~K. (c)
  Schematic illustration of the energy balance for polaron formation in
  the presence of lattice distortions.}
 \end{figure}
We then analyzed the thermal properties of the Mu$^+_i$-bound small
polaron by examining the temperature dependence of $\mu^+$SR spectra.
Figure~\ref{fig2}(a) shows the temperature evolution of FT-$\mu^+$SR 
spectra in a TF of 0.5~T applied along the [001] direction. As
temperature increased, a decrease in the paramagnetic fraction and an increase in the
diamagnetic one were observed, suggesting that thermal dissociation of
the Mu$^+_i$-bound small polaron occurred.
To obtain a more detailed analysis of the dissociation process, we 
performed fits in the time domain to an exponentially-damped cosine function using data 
points after 3~$\mu$s, where the paramagnetic signals were mostly damped.
Figure~\ref{fig2}(b) displays the temperature dependence of the diamagnetic 
fraction obtained by the fits, which is qualitatively consistent
with that reported in Ref.~[\cite{salman14}] under a low TF of 10~mT.
We used the ionization model from Ref.~[\cite{salman14}] 
to fit our data and obtained a characteristic energy $E_a$ of 30(3)~meV.
The $E_a$ in the ionization model usually represents the depth of a
shallow Mu$^0$ donor level measured from the conduction band
minimum; however, in this case, it should be interpreted as the activation
energy for the Mu$^+_i$-polaron dissociation. The electron separated
from the Mu$^+_i$ donor is expected to serve as a charge carrier
in the form of either a delocalized conduction-band electron or a hopping small
polaron.
The small $E_a$, 
comparable in magnitude with the characteristic thermal energy at room
temperature, 
indicates that the interstitial doping of Mu (or H) into SrTiO$_3$
leads to $n$-type conductivity, which is consistent with hydrogen doping
studies.\cite{nakayama18}

The fate of the electron released from the Mu$^+_i$-polaron complex depends
on the nature of excess electrons in the perfect SrTiO$_3$ lattice.
Some computational studies suggest that excess electrons in SrTiO$_3$
tend to delocalize rather than create self-trapped small polarons unless
Coulomb and strain fields due to ionized dopants are
present.\cite{janotti14,hao15}
This is in sharp contrast to the situation in rutile TiO$_2$, where the stability of
self-trapped small polarons has been established from both theoretical and
experimental viewpoints.\cite{setvin14,janotti13,yang13}
Thus, we assume that the electron released from the Mu$^+_i$-polaron
complex is immediately transferred to the conduction band in SrTiO$_3$.
This assumption is also consistent with the coherent delocalized state
detected by PES in $n$-type SrTiO$_3$.\cite{ishida08,dangelo12}
There may seem to be a disagreement between the small $E_a$ and the highly localized nature of
the small polaron, which usually forms a deep in-gap state.\cite{janotti14,hao15,setvin14} 
This could be explained by considering the energy balance for polaron
formation in the presence of a lattice distortion due to the localized
electron and ionized dopant,\cite{hao15,setvin14,janotti13} as
illustrated in Fig.~\ref{fig2}(c).
The electronic energy $E_{el}$, associated with the deep single-particle
state at $\sim1$~eV below the conduction band minimum,
is the quantity measured by PES or scanning
tunneling spectroscopy within the time scale where lattice atoms are
considered to be frozen (vertical excitation). In addition, the polaron formation energy
$E_{pol}$ is defined as
the total energy difference between the polaronic and delocalized states
in fully relaxed lattices.
These energies are connected through the relation $E_{pol}=E_{el}-E_{st}$ with the
strain energy $E_{st}$ required to distort the lattice.
In the thermal dissociation process associated with the $E_a$ obtained from
$\mu^+$SR,
the polaronic electron may be transferred to the conduction
band via a minimum energy path in Fig.~\ref{fig2}(c) with an energy barrier
comparable to $E_{pol}$.
The value for $E_{pol}$ was estimated for Nb-doped SrTiO$_3$ to be in the range of
several tens of meV by first-principles calculations.\cite{hao15}
This is comparable in magnitude with our estimation for Mu(H)-doped SrTiO$_3$
in the dilute doping limit. 
When $E_{pol}\ll E_{el}\sim E_{st}$, strain and Coulomb
fields induced by ionized dopants must play a crucial role in
stabilizing the small polaron ground state.

There are some important discrepancies between this $\mu^+$SR
study and the preceding work by Salman {\it et al.}\cite{salman14}
They determined the principal values of the hyperfine coupling tensor
$\bm{\mathsf{A^{\prime}}}$ as 1.4(1), 6.7(1), and 11.5(1) MHz from
$\mu^+$SR frequencies observed at 25~K in ZF. Because the dipolar 
part in $\bm{\mathsf{A^{\prime}}}$ should be traceless, the coefficient of the
isotropic part is estimated as
Tr$[\bm{\mathsf{A^{\prime}}}]/3=6.5(1)$~MHz,
which is comparable in magnitude with the largest principal value.
This implies that the muon is located in an extended electron cloud as in
the case of a hydrogenic shallow donor. Contrastingly, 
we concluded from the $A_{dip}\gg A_{iso}$ relation at 1.7~K that
the electron spin density is dominantly distributed on a
Ti site adjacent to an ionized Mu$^+_i$. This discrepancy may 
be partly due to the fact that the hyperfine parameters were
investigated at different temperatures. The splitting of the
paramagnetic lines seems mostly symmetric at 25~K, while it is
undoubtedly asymmetric at 1.7~K (Fig.~\ref{fig2}(a)).
This change is probably attributed to the activation of {\it local}
Mu$^+$ motion among four quasi-equivalent sites around the nearest
Ti$^{3+}$-O$^{2-}$ bond in our model. In this situation, the component
of the hyperfine field perpendicular to $\bm B$ is averaged to zero and
the remaining parallel component causes the symmetric splitting.
In Ref.~[\cite{salman14}], TF dependences of
paramagnetic frequencies at 25~K below 12~mT were analyzed with
$\bm{\mathsf{A^{\prime}}}$ expressed in the laboratory frame using two
sets of Euler angles. The high-field extension of paramagnetic frequency
shifts calculated with these parameters is shown in Fig.~\ref{fig1}(a) 
(dashed lines). These parameters obviously fail to explain our experimental
results at 1.7~K.
Another difference is the interpretation of the small $E_a$.
It was previously interpreted as an ionization energy of a shallow
hydrogen-like Mu$^0$,\cite{salman14} while we assigned it to the
effective binding energy of a polaronic electron stabilized near
Mu$^+_i$.
In our model, a single-particle level associated with the
well-localized electron is expected to lie deep in the band gap as 
observed in other electron-doped SrTiO$_3$
systems.\cite{ishida08,dangelo12}

In conclusion, we observed a spectroscopic signature of a Mu$^+_i$-bound
small polaron in bulk SrTiO$_3$, which easily dissociates at moderate
temperatures ($E_a=30(3)$~meV).
In the dilute doping limit, H$_i$ is also expected 
to form such a polaronic center according to the hydrogen-Mu analogy.
The direct observation of a deep in-gap state by PES in the bulk
crystal of H$_i$-doped SrTiO$_3$ would be crucial to identify the
H$_i^+$-bound small polaron state corresponding to the muonic one.

%
%
%

\begin{acknowledgments}
We thank the staff of the PSI muon facility for technical
 assistance and K.~Nishiyama and K.~Fukutani for helpful
 discussions. This work was partially supported by Grant-in-Aid for
 Young Scientists (No.16K17544) from Japan Society for the Promotion of Science.
\end{acknowledgments}


\begin{thebibliography}{99}
\bibitem{schooley64} J.~F.~Schooley, W.~R.~Hosler, and M.~L.~Cohen, 
        Phys.~Rev.~Lett. {\bf 12}, 474 (1964).
\bibitem{rice14} W.~D.~Rice, P.~Ambwani, M.~Bombeck, J.~D.~Thompson,
        G.~Haugstad, C.~Leighton, and S.~A.~Crooker,  
        Nat.~Mater. {\bf 13}, 481 (2014).
\bibitem{ohtomo04} A.~Ohtomo and H.~Y.~Hwang,  
        Nature {\bf 427}, 423 (2004).
\bibitem{yoshimatsu08} K.~Yoshimatsu, R.~Yasuhara, H.~Kumigashira, and
        M.~Oshima, 
        Phys.~Rev.~Lett. {\bf 101}, 026802 (2008).
\bibitem{mechelen08} J.~L.~M.~van~Mechelen, D.~van~der~Marel,
	C.~Grimaldi, A.~B.~Kuzmenko, N.~P.~Armitage, N.~Reyren,
	H.~Hagemann, and I.~I.~Mazin, 
	Phys.~Rev.~Lett. {\bf 100}, 226403 (2008).
\bibitem{devreese10} J.~T.~Devreese, S.~N.~Klimin,
	J.~L.~M.~van~Mechelen, and D.~van~der~Marel, 
	Phys.~Rev.~B {\bf 81}, 125119 (2010).


\bibitem{ishida08} Y.~Ishida, R.~Eguchi, M.~Matsunami, K.~Horiba,
	M.~Taguchi, A.~Chainani, Y.~Senba, H.~Ohashi, H.~Ohta, and
	S.~Shin, 
	Phys.~Rev.~Lett. {\bf 100}, 056401 (2008).
\bibitem{lee11} M.~Lee, J.~R.~Williams, S.~Zhang, C.~D.~Frisbie, and
	D.~Goldhaber-Gordon, 
	Phys.~Rev.~Lett. {\bf 107}, 256601 (2011).


\bibitem{janotti14} A.~Janotti, J.~B.~Varley, M.~Choi, and
	C.~G.~Van~de~Walle, 
	Phys.~Rev.~B {\bf 90}, 085202 (2014).
\bibitem{hao15} X.~Hao, Z.~Wang, M.~Schmid, U.~Diebold, and
	C.~Franchini, 
	Phys.~Rev.~B {\bf 91}, 085204 (2015).
\bibitem{iwazaki14} Y.~Iwazaki, Y.~Gohda, and S.~Tsuneyuki,
	APL Mater. {\bf 2}, 012103 (2014). 
	
\bibitem{nakayama18} R.~Nakayama, M.~Maesato, T.~Yamamoto, H.~Kageyama,
	T.~Terashima, and H.~Kitagawa, 
	Chem.~Commun. {\bf 54}, 12439 (2018).
\bibitem{dangelo12} M.~D'Angelo, R.~Yukawa, K.~Ozawa, S.~Yamamoto,
	T.~Hirahara, S.~Hasegawa, M.~G.~Silly, F.~Sirotti, and
	I.~Matsuda,  
	Phys.~Rev.~Lett. {\bf 108}, 116802 (2012).
\bibitem{kapphan80} S.~Kapphan, J.~Koppitz, and G.~Weber, 
	Ferroelectrics {\bf 25}, 585 (1980).
\bibitem{weber86} G.~Weber, S.~Kapphan, and M.~W\"{o}hlecke
	Phys.~Rev.~B {\bf 34}, 8406 (1986).

\bibitem{klauer94} S.~Klauer and M. W\"{o}hlecke, 
	Phys.~Rev.~B {\bf 49}, 158 (1994).


\bibitem{spencer83} D.~P.~Spencer, D.~G.~Fleming, and J.~H.~Brewer,  
	Hyperfine Interactions {\bf 17-19}, 567 (1984).
\bibitem{salman14} Z.~Salman, T.~Prokscha, A.~Amato, E.~Morenzoni,
	R.~Scheuermann, K.~Sedlak, and A.~Suter,  
	Phys.~Rev.~Lett. {\bf 113}, 156801 (2014).


\bibitem{cox06} S.~F.~J.~Cox, J.~L.~Gavartin, J.~S.~Lord,
	S.~P.~Cottrell, J.~M.~Gil, H.~V.~Alberto, J.~Piroto~Duarte,
	R.~C.~Vil\~ao, N.~Ayres~de~Campos, D.~J.~Keeble, E.~A.~Davis,
	M.~Charlton, and D.~P.~van~der~Werf, 
	J.~Phys.: Condens.~Matter {\bf 18}, 1079 (2006).
\bibitem{ito13} T.~U.~Ito, W.~Higemoto, T.~D.~Matsuda, A.~Koda,
	K.~Shimomura,  Appl.~Phys.~Lett. {\bf 103}, 042905 (2013).
\bibitem{ito17} T.~U.~Ito, A.~Koda, K.~Shimomura, W.~Higemoto,
	T.~Matsuzaki, Y.~Kobayashi, and H.~Kageyama, 
	Phys.~Rev.~B {\bf 95}, 020301(R) (2017).

	
\bibitem{vilao15} R.~C.~Vil\~ao, R.~B.~L.~Vieira, H.~V.~Alberto,
	J.~M.~Gil, A.~Weidinger, R.~L.~Lichti, B.~B.~Baker,
	P.~W.~Mengyan, and J.~S.~Lord, 
	Phys.~Rev.~B {\bf 92}, 081202(R) (2015).
\bibitem{shimomura15} K.~Shimomura, R.~Kadono, A.~Koda, K.~Nishiyama,
	and M.~Mihara,  
	Phys.~Rev.~B {\bf 92}, 075203 (2015).
	

\bibitem{amato17} A.~Amato, H.~Luetkens, K.~Sedlak, A.~Stoykov,
	R.~Scheuermann, M.~Elender, A.~Raselli, and D.~Graf, 
	Rev.~Sci.~Inst. {\bf 88}, 093301 (2017).



\bibitem{setvin14} M.~Setvin, C.~Franchini, X.~Hao, M.~Schmid,
	A.~Janotti, M.~Kaltak, C.~G.~Van~de~Walle, G.~Kresse, and
	U.~Diebold, 
	Phys.~Rev.~Lett. {\bf 113}, 086402 (2014).

\bibitem{janotti13} A.~Janotti, C.~Franchini, J.~B.~Varley, G.~Kresse,
	and C.~G.~Van~de~Walle, 
	Phys.~Status~Solidi~PRL {\bf 7}, 199 (2013).

\bibitem{yang13} S.~Yang, A.~T.~Brant, N.~C.~Giles, and
	L.~E.~Halliburton, 
	Phys.~Rev.~B {\bf 87}, 125201 (2013).


	
	 


 \end{thebibliography}
\end{document}